\documentclass[proof]{pasj00}

\usepackage{lscape}

\begin{document}

\title{The KaVA and KVN Pulsar Project}

 \author{Richard \textsc{Dodson}\altaffilmark{1,2} 
 \email{rdodson@kasi.re.kr},
   Chunglee \textsc{Kim}\altaffilmark{3},
   Bongwon \textsc{Sohn}\altaffilmark{1},
   Mar\'{\i}a J. \textsc{Rioja}\altaffilmark{1,2,4},
   Taehyun \textsc{Jung}\altaffilmark{1},
   Andrew \textsc{Seymour}\altaffilmark{5},
   Wasim \textsc{Raja}\altaffilmark{6},
}
 \altaffiltext{1}{Korea Astronomy and Space Science Institute (KASI), Daedeokdae-ro 776, Yuseong-gu, Daejeon 305-348, Korea}
 \altaffiltext{2}{International Centre for Radio Astronomy Research, M468,The University of Western Australia, 35 Stirling Hwy, Crawley, Western Australia, 6009}
 \altaffiltext{3}{Astronomy Program, Dept.\ of Physics and Astronomy, Seoul National University, 1 Gwanak-ro, Gwanak-gu, Seoul 151-742, Korea}
 \altaffiltext{4}{Observatorio Astron\'omico Nacional (IGN), Alfonso XII, 3 y 5, 28014 Madrid, Spain}
 \altaffiltext{5}{Department of Physics and Astronomy, West Virginia University, PO Box 6315, Morgantown, WV 26506, USA}
 \altaffiltext{6}{Raman Research Institute, Bangalore, 560080 Karnataka India}
%%% end:list of authors

%Do NOT move this preamble from here!

\maketitle

\begin{abstract}

We present our work towards using the Korean VLBI (Very Long Baseline Interferometer) Network (KVN) and VLBI Exploration of Radio Astronomy (VERA) arrays combined into the KVN and VERA Array (KaVA) for observations of radio pulsars at high frequencies ($\simeq$22-GHz). Pulsar astronomy is generally focused at frequencies approximately 0.3 to several GHz and pulsars are usually discovered and monitored with large, single-dish, radio telescopes. For most pulsars, reduced radio flux is expected at high frequencies due to their steep spectrum, but there are exceptions where high frequency observations can be useful. Moreover, some pulsars are observable at high frequencies only, such as those close to the Galactic Center. The discoveries of a radio-bright magnetar and a few dozen extended {\it Chandra} sources within 15\,arc-minute of the Galactic Center provide strong motivations to make use of the KaVA frequency band for searching pulsars in this region. Here, we describe the science targets and report progresses made from the KVN test observations for known pulsars. We then discuss why KaVA pulsar observations are compelling.

\end{abstract}

\section{Introduction}

\subsection{Pulsar Targets}

Radio pulsars (PSR) are the compact remnants of massive stars that have undergone catastrophic collapses into neutron stars. They are highly magnetized and emit a narrow beam of electromagnetic radiation. Emission from pulsars is observed as pulses as the beam is swept across an observer's line-of-sight. Pulsar emission can run from tens of MHz in the radio bands, through optical, X-ray, and in some cases, gamma rays (e.g. the simultaneous observations of Radio and X-ray pulses from the Vela Pulsar \citep{lommen_07}). Radio pulsars are considered as one of nature's most stable clocks. There are roughly 2000 radio pulsars known in our Galaxy including globular clusters and Small/Large Magellanic clouds as listed in the ATNF pulsar catalogue\footnote{http://www.atnf.csiro.au/people/pulsar/psrcat} \citep{man_05}. Galactic pulsar populations are diverse, and usually they are classified by two parameters (spin period and period derivative). The majority of known pulsars, sometimes called `normal', have typical spin periods of $\sim$1 s and period derivatives of $\sim10^{-15}$ ss$^{-1}$. Most pulsars are solitary, but some are bound in binaries with other objects. A few dozen pulsars have either white dwarfs or neutron stars as their companions. 
%Multi-frequency observations are used to understand the pulsar's emission mechanism and magnetosphere. %Pulsar VLBI 

In the context of KaVA pulsar observation, there are two groups of pulsars of interest;
one is a theoretically predicted population: fast pulsars close to the Galactic Center. The other is the observationally known pulsars and pulsar candidates in the central region of our Galaxy. Such fast-spinning pulsars in the Galactic Center would be useful probes for studying the gravitational field and environment around the central black hole in our Galaxy (e.g. \cite{pfahl_04, liu_12}). 
%It is advantageous to have a fast-spinning pulsars (e.g. spin periods a few tens of milliseconds or up to 1 second) and to measure their pulse arrival times with good accuracy. 
Deviations from expected pulse arrival times can be interpreted as the effects of the environment and, most interestingly, the central black hole. Detecting pulsars close to the central black hole is considered to be enormously challenging due to the severe observational difficulties. At lower frequencies, that is below several GHz, pulse broadening due to scattering from free electrons is severe around the Galactic Center. High-frequency observations well above 10-GHz are required to detect pulsars in the central region of our Galaxy (e.g. \cite{cordes_97, macquart_gbt, wharton_12}). Even higher frequencies are required in order to detect fast-spinning pulsars, e.g. for spin periods less than 20-ms.

So far, there have been several international efforts to find pulsars toward the Galactic Center at up to $\sim$18-GHz, but no survey has been done above 20-GHz. \citet{johnston_06} and \citet{deneva_09} discovered a few pulsars, with separations of arc-minutes to degrees from Sgr\,A*. There are other surveys toward the Galactic Center, but these produced no detections (see \citet{wharton_12} and \citet{chennamangalam_14} for more details of previous efforts). As one example, \citet{macquart_gbt} performed a deep pulsar search within the central parsec around Sgr\,A* at 15-GHz with the Green Bank telescope, which is one of the highest frequencies attempted to-date. They report null detection with $10\sigma$ threshold of 10\,$\mu$Jy, but suggest that they would expect $\sim$90 normal pulsars to reside within that area. Recently, \citet{chennamangalam_14} suggested that the number of detectable pulsars in the Galactic center region is $\sim$200, considering recent estimates made for the pulse broadening \citep{spitler_14} and angular broadening \citep{bower_14} in this region. Considering the uncertainties in the environment and neutron star population suggested by previous works, 1--100 detectable radio pulsars would be expected within a few parsecs about Sgr A*. In addition to single pulsars, theoretical predictions suggest that wide, eccentric binaries consisting of a fast-spinning pulsar and stellar-mass black hole, can be formed by stellar interactions and reside within a parsec of Sgr A* \citep{fgl_11}.

The normal pulsar spectrum is steep; the latest work suggests that the data can be fitted with a mean spectral index of $\sim$-1.4 \citep{bates_13}. However, roughly a dozen pulsars are reported to have flat or flatter spectrum than normal pulsars. For example, \citet{kramer_97} detected four pulsars (B0329+54 \citep{cp68}, B0355+54 \citep{mth72}, B1929+10 \citep{lvw68}, and B2021+51 \citep{dl70} at 15, 23, and 43-GHz. Further high-frequency pulsar observations have been done by various groups e.g. \citet{kramer_96, lohmer_08, keith_11}. %
Some of the flat-spectrum pulsars are categorized as magnetars, i.e.\ neutron stars with magnetically powered emission, which were first theoretically explained by \citet{thompson_95}. There are roughly 30 magnetars known to-date (see the McGill Magnetar Catalog\footnote{http://www.physics.mcgill.ca/~pulsar/magnetar/main.html}, \cite{olausen_14}), including the first discovered radio-bright magnetar XTE J1810-197 \citep{camilo_06}. At frequencies higher than 20-GHz, this magnetar is the brightest among known pulsars.

Recently such a magnetar has been discovered in the Galactic Center region. 
In April 2013 a 3.76sec pulsating transient source 
was detected, originally by the Swift and NuStar X-ray satellites. Shortly after the announcement of the discovery (AstroTel $\#5020$ \citep{ATel_5020}), the source was confirmed in radio and identified as a `magnetar' (see \#5064, and others). This magnetar is unique due to its location, as it is only 3$^{\prime \prime}$ away from Sgr\,A* on the sky. The pulse-averaged radio flux density estimated by the 100-m Efflesberg observations at 18.95-GHz is $\sim$0.2 mJy (see \cite{ATel_5058}). The pulsar emission is concentrated into a short duration pulse of $\sim$2\% of the total period \citep{eatough_13}, so the implied peak flux-density is $\sim$10mJy. The emission has an almost flat spectrum. Based on the angular distance from Sgr\,A* and the estimated dispersion measure of $\sim$1700 pc cm$^{-3}$, this magnetar must indeed be associated with Sgr\,A* \citep{mori_13,shannon_13}.

In order to beat the scatter broadening observations at high-frequencies are the key to detect fast-spinning pulsars near Sgr A*. However, because of the pulsar spectral index, the fluxes at these frequencies are much lower and highly sensitive observations with the maximum collecting area and bandwidth are required. This is the opportunity that our observations with KVN and VERA (KaVA) aim to fullfil.

\subsection{Advantages of observing with KaVA}

Korean and Japanese VLBI radio astronomers are currently operating their respective VLBI arrays in conjunction as the KaVA network, for 50\% of their operating time. The KVN consists of three antennas on the Korean peninsular, hosted at the Universities of Yonsei, Ulsan and Tamna \citep{kvn}. The maximum baselines for KVN are about 500\,km; such baselines are short for VLBI arrays. The VERA array consists of four antennas which provide the maximum baselines across Japan, as they are located at Mizusawa, Ishigaki, Iriki and Ogasawara \citep{vera}. The maximum baselines for VERA are about 2300\,km, but the minimum is 1000\,km; VERA alone has no short baselines. By operating the two arrays in conjunction the uv-coverage for KaVA, compared to that for KVN and VERA independently, is greatly improved. 

Pulsar timing observations are normally attempted with large single dish telescopes. The antennas of KaVA are modest-sized ($\sim$20-m) telescopes, but if the dishes are added coherently they sum to the equivalent of a 54-m antenna. This is sufficient to make KaVA a competitive, as well as an extremely innovative, instrument. Furthermore VLBI measurements at 22-GHz can provide unprecedented precision to measure the location of the pulsars, based on phase referencing. Most pulsar phase referencing is done around 20-cm, a wavelength nearly fifteen times longer than those we plan to use \citep{brisken_02,chatterjee_09}. The resolution, and therefore the astrometric accuracy, scales with the frequency. The penalty of course is the reduced signal to noise (SNR) which also reduces astrometric accuracy. Therefore, for some pulsars, we can expect better astrometry at these higher frequencies than for the lower frequencies conventionally used in pulsar astronomy. 
When performing pulsar VLBI observations it is standard practise to correlate the pulsar data with the output divided into  subsets, or bins, of the pulsar period. When the pulsar ephemerises are well known ahead of correlation, only the data which holds the pulsar signal needs to be retained, so called pulsar-gated data \citep{gating}. When the pulsar period is reasonably known, but the pulsar phase is not, all the subsets need to be retained and analysed, which is known as pulsar-binned data. In both cases, the data can be referred to as a pulsar-phase resolved VLBI dataset. 
However, pulsar phase resolved VLBI only works when the pulsar timing solution is accurately known, i.e. when the timing solution does not change more than a phase bin over the span of the experiment. Thus poorly known or unknown pulsars can not be directly imaged. The largest part of our project is to discover completely unknown pulsars and therefore we need to be able to generate timing `filterbank' outputs from the VLBI datasets. Pulsar timing analysis works on data which has been averaged in time to, typically, a fraction of a millisecond (ms) and averaged in frequency to, typically, a fraction of a MHz. For a single dish antenna, the data that are averaged is the receiver power. If we are to use the well-established pulsar searching methods we need to provide data compatible with these tools. Therefore, we developed the procedures required to produce a fine time resolution autocorrelation spectrum from the VLBI data, and to extend the approach to visibility data. 
% I don't understand this sentence xxxx
%Here, we note that we use `pulsar timing' in comparison of `pulsar imaging' in the context of VLBI observations. The former is simply means the periodicity search in time (or frequency) domain.

\section{Project Plan}

\subsection{Science Goals}

We believe that KaVA can significantly contribute to improved astrometry of known pulsars, especially those with flatter spectrum. Moreover, the expected KaVA sensitivity with the methods we describe in the next section will allow us to perform deep pulsar searches in the Galactic Center region. The KaVA pulsar project has two strands: \underline{Imaging} to provide VLBI positions with pulsar phase resolved imaging, and \underline{Timing} to search for pulsars at 22-GHz. The latter will be done with: incoherently summed single dish filterbank outputs, coherent phased-array finely sampled time-domain outputs and coherent time-domain filterbank outputs based on the Bi-spectrum correlated data-stream (that is, to use the closure product). 
In preparation for this goal, we are observing a selection of interesting known pulsars who will deliver important science and demonstrate our methods. 
% We performed two sets of observations in May and November 2013 with the KVN. We have had our first KaVA observations in January 2014.

The scientific themes are:

%\begin{itemize}
   $\bullet$ {\bf Pulsar VLBI observation of known pulsars at high frequencies:} 
%During the past decade, Radio Astronomy has begun to study transient science (e.g. The ASKAP-CRAFT project (Dodson et al. 2009)). 
There are more than a dozen pulsars that show giant radio pulses or intermittent pulsed emission, and these often have a relatively flat radio spectrum. All of these are in general worthy of  observation in the KaVA frequency bands. We have selected for preliminary observations an interesting pulsar, B0355+54, which is bright at 22-GHz. 
Long-term monitoring of known pulsars with flatter spectra at 22-GHz (and above) will shed light on the understanding of the pulsar emission mechanism and provide a bridge between lower-frequency radio observation and those with Atacama Large Millimeter Array (ALMA).   
%
%  $\bullet$ {\bf Precision pulsar astrometry for PSR\,J1745-2900:} 
Of particular interest will be observations of the unique magnetar PSR\,J1745-2900 \citep{mori_13,eatough_13,shannon_13}, which is only 3$^{\prime \prime}$ distant from Sgr\,A*. Phase referencing with Sgr\,A* can provide a phase referenced astrometric position for this magnetar. In order to monitor its movement in the gravitational field of the Galactic Center, it is essential to constrain the pulsar location as precisely as possible with VLBI astrometry. KaVA is capable of providing this and will provide 4 times the resolution of the KVN alone. 
Because the pulsar is so close to Sgr\,A*, the observational requirements are identical to that of the Sgr\,A* monitoring project, and we will recorrelate the data from that project, in DiFX, to form a pulsar phase resolved dataset.

$\bullet$ {\bf Detection of High-DM pulsars and Mapping the spacetime around Sgr\,A*:} 
As mentioned in the introduction, several authors have made theoretical and empirical studies that predict that there are numerous high Dispersion Measure (DM) pulsars to be discovered close to Sgr\,A* \citep{fgl_11, wharton_12, chennamangalam_14}.  These will have significant scientific impact \citep{liu_12}. %
Observationally there are a number  of peculiar {\it Chandra} sources of X-ray emission in this region. These have been proposed to be Pulsar Wind Nebulae (PWN) and therefore should host pulsars  \citep{lu_08, muno_08, johnston_09} and these would represent some of the pulsars predicted by the simulations. 
%There must be other such sources without X-ray detections. 
We will undertake deep pulsar timing searches with KaVA of the Galactic Center region. We have three approaches for this challenge: i) incoherent timing searches of the primary beam, ii) coherent timing searches on selected targeted positions focused on the Chandra X-ray sources and iii) coherent bi-spectrum timing searches of the correlator Field of View. We expand on these below:

\indent i) Incoherent timing searches of the sum of the autocorrelations from the KaVA array gives us the same sensitivity as a 34m dish. This is useful and simple to perform. \\
\indent ii) Timing searches of coherent phased-array beams from the array are significantly more sensitive, the equivalent to a 54m dish, but requires accurate target positions. There are 5 {\it Chandra} sources with filament-like morphology within the KaVA beam at Sgr\,A*.  We have selected these as our initial target positions, along with that of PSR\,J1745-2900, for our phased array analysis.\\
\indent iii) Bispectrum timing searching is a new method which has the capacity to deliver coherent sensitivity across the whole primary beam, and was proposed by \citet{law_12}. This is a new route for the Square Kilometer Array (SKA) pulsar observations, and has great - but undemonstrated - potential. We will search the whole primary beam of the KaVA antennas for transient and pulsar sources.\\

\subsection{Methods}
\label{sec:meth}

To achieve the above goals we are planning to observe pulsars in four modes:
\begin{enumerate} 
\item Pulsar Phase-resolved VLBI-imaging of the cross correlations, for imaging.
\item Phased-array summed cross-correlations, for timing.
\item Multi-telescope summed Single Dish autocorrelations, for timing.
\item Bispectrum summed closure triangles, for timing.
\end{enumerate}

The first is widely used and easy to achieve, with the DiFX \citep{difx} correlator for example. One requires ahead of time both a reasonably accurate pulsar position (within the correlator Field of View, which is typically some arc-seconds) and a precise pulsar timing solution. The pulsar must not drift with respect to the provided timing parameters over the whole observing session which, if one was to hold to a millisecond over six hours, requires accuracies of 1 part in $\sim$22\,million. Therefore, concurrent timing solutions are highly desirable. When observed with a phase reference source, pulsar phase resolved VLBI will provide accurate astrometric positions for the pulsar as a function of observing epoch. Therefore, this is the observing mode we will focus on. There are some cases when the pulsar and the calibrator can be simultaneously observed within a single antenna pointing. We can perform such `in-beam' calibration when we are searching around Sgr\,A*, as that source would provide the required calibration for the imaging of the other fields in the primary beam.

The second method sums the cross correlations in phase, the so called phased array mode. The output is a high time resolution (in principle down to the Nyquist sampling rate) dataset, coherently averaged in frequency and time for processing with the usual pulsar timing analysis tools. The phased array mode requires that the time-variable calibration values for the array are known and that the pulsar is within the synthesised beam of the array. That in turn requires that there are calibrators included in the observations and the solutions from these can be applied during the correlation. As such, it has the same requirements as in the first case, for the astrometric VLBI imaging.  Furthermore, this correlation will only be valid for a very small area on the sky, that of the synthesised beam size, which is the order of about a milli-arcsecond (mas). Therefore, this approach is only suitable when one knows where the target is, but not the pulsar period. As such, it is ideal for searches of pulsar candidate sites. The JIVE software correlator (SFXC, \cite{kettenis_SFXC}) can provide the required phase array outputs, which can then be fed into the standard pulsar searching tools. We have been using {\tt PRESTO} \citep{presto} which finds the period of the pulsed emission by searching over a range of allowed parameters, such as period and derivatives, DM or binary orbit elements. 

The third is conceptually trivial and we have developed software to facilitate it. Forming high time resolution autocorrelations from wide bandwidth observations is normally done with dedicated hardware, but we have added software to the DiFX code tree to provide this from recorded VLBI data. The filterbank is formed by doing a Fourier transform of the samples from each data-stream with the number of channels required to provide the requested spectral frequency resolution and summing the amplitudes up to the temporal integration period requested. Nearly all VLBI recording modes are supported by the DiFX libraries, including the major ones used in Japan and Korea, and this software is now included as part of the DiFX tree. The autocorrelations formed in the software filterbank are sensitive to any source of pulsation which is in the field of view (FoV) of the single antennas; that is $\sim$2$^{\prime}$ at 22-GHz. The output can be fed into the standard pulsar searching tools such as {\tt PRESTO}. Aligning datasets from multiple telescopes requires each data set to be initially processed separately to correct for the different tropocenters for each antenna, but after this step is done they can be treated as additional IFs in the normal dataprocessing chain and combined incoherently. For this mode, the preferred observing mode would be a single long timespan track of the target, which is not compatible with the requirements for the two phase referenced observing modes mentioned above. However {\tt PRESTO} supports flagging tables, so it is trivial to flag the off-source data before feeding into the timing search. The phase referencing temporal sampling, typically of a few minutes cycle time, will have a very limited effect on searches for pulsars with periods of milliseconds to seconds. 
 
The fourth method is to form the Bispectrum products \citep{law_12} as an additional output of the correlation. The bispectrum is also known as the closure triangle as it is formed by the product of $V_{ij} \times V_{jk} \times V_{ki}$ where $V_{ij}$ is the complex visibility function on the baseline between antennas {\em i} and {\em j}. The product of the visibility of the baseline between antenna one and two, that between antenna two and three and that between three and one, leads to the cancellation of all antenna-based residual phase terms, such as receiver delays and phases and also the source position relative to the center of the field. We have implemented the generation of this product within the DiFX correlator and we are able to search the output data stream in the same manner as conventional pulsar searching. The  Bispectrum product is equally sensitive to the whole field of view of the correlator (ignoring primary beam effects).  The correlator field of view is limited by two effects, the temporal and bandwidth smearing. 
As the temporal smearing for the Bispectrum is limited by the sub-integration time sampling, which is necessarily very small as one is searching for fast pulsar or transient signals, only the bandwidth smearing contributes to the limitations of the correlator field of view. To observe the whole field of view of the individual antennas, the ratio of the channel width to the observing frequency needs to be small in comparison to the ratio of the dish size to the baseline length. In the case of 20\,m dishes and 3000\,km baselines the individual channel width should be significantly less than 0.1-MHz at an observing frequency of 22-GHz. If the frequency sampling is cruder than this the angular size of the correlator field of view is linearly reduced to the center of the antenna pointing. As for the previous case, a single pointing of the target field of view would be preferred, but the off-source data can be easily flagged. 
% KVN ~OK with 16MHz, KaVA not

Therefore, the target cases for each of these methods are:
\begin{enumerate} 
\item to image and measure the precise positions of known pulsars with known periods. %Phase resolved VLBI imaging
\item to coherently search the sites of pulsar candidates for a range of periods,  to discover new pulsars for which we have possible accurate positions.  %Phase array summed cross correlations
\item to incoherently search for all unknown pulsars with unknown periods within the primary beam FoV, to discover new pulsars using conventional methods. %Multi-telescope summed Single Dish autocorrelations
\item to coherently search for all unknown pulsars with unknown periods and transients within the correlator FoV, to discover new pulsars using new and innovative methods.  %Bispectrum summed closure triangles
\end{enumerate}

\subsection{Sensitivity Analysis}

\begin{figure}[h]
    \includegraphics[width=0.99\textwidth,angle=0]{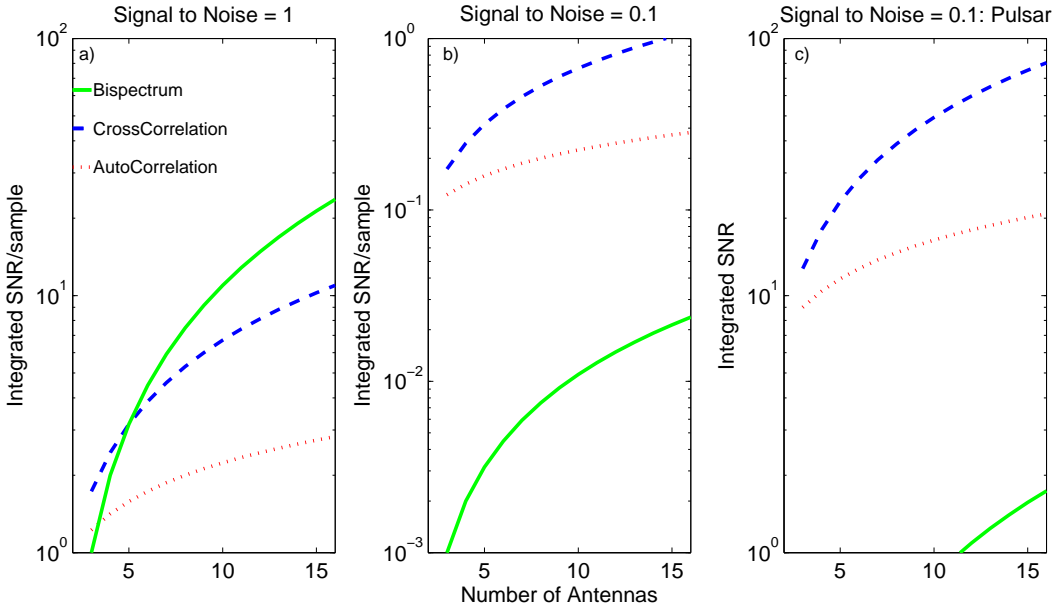}
  \caption{Signal to Noise Ratio calculations for the indicative Signal to Noise cases: SNR per sample of 1.0 (left) and 0.1 (middle). The right-hand plot is for a SNR of 0.1 per sample, but integrated across a whole experiment. For the strong signal case (left) the bispectrum has the best response for N$_{\rm ant}>$4. This would be true even if the signal is made up of a single sample, i.e. for a transient or a single strong pulsar signal. However for the weak SNR case (middle) the bispectrum has far the worst sensitivity and furthermore no pulse is detectable in a single sample for any method. Nevertheless integrating many weak pulsations together, 5400 in this case representing 6 hours of observing a 4-second period pulsar, would produce a detectable signal for both the methods based on the incoherent sum of autocorrelations and on the coherent sum of phased array correlations but not the bispectrum method.}
\label{fig:snr}
\end{figure}

The sensitivity of each of the cross correlation, autocorrelation and bispectrum approaches have very different dependencies. The first two methods, that of imaging and the phased array, have identical sensitivities as each pixel of the image is made up of the phased sum of the cross correlations. The discussion of these two is combined in the following section. However, the autocorrelation and the bispectrum do have different sensitivities, as listed below:

\begin{itemize} 
%\item $S \sqrt{N_{\rm obs} C^{N_{\rm A}}_2}$. %Phase resolved VLBI imaging
\item Cross Correlations: $S \sqrt{N_{\rm obs} C^{N_{\rm A}}_2}$  %Phase array summed cross correlations
\item Auto Correlations: $\frac{S} {\sqrt{2}} \sqrt{N_{\rm obs} C^{N_{\rm A}}_1}$  %Multi-telescope summed Single Dish autocorrelations
\item Bispectrum: $S^3 \sqrt{N_{\rm obs} C^{N_{\rm A}}_3}$  %Bispectrum summed closure triangles
\end{itemize}

where $N_{\rm obs}$ is the number of observed data points, $N_{\rm A}$ is the number of antennas, $C^{N_{\rm A}}_r$ is the number of combinations of $r$ correlations (i.e. r=1 for Autocorrelation, 2 for Cross correlations and 3 for closure triangles) from the $N_{\rm A}$ antennas and $S$ is the SNR in one visibility product on one baseline. Note that $S$ is the SNR  per frequency channel per time sample rather than that for the entire dataset. For the linear cases (the first two) these are equivalent; the losses in sensitivity in individual data samples, caused by the finer sampling, are recovered by having more samples. However for the bispectrum case, oversampling by a factor $\epsilon$ will lead to a loss of SNR of $\epsilon^{3/2}$ of which only $\epsilon^{1/2}$ will be recovered by the increased number of samples, $N_{\rm obs} \times \epsilon$. Figure \ref{fig:snr} explores the consequences of this effect. 

Figure \ref{fig:snr}a shows the SNR summed  across antennas, as a function of the number of antennas, for cases where the SNR per sample is 1; that is the signal in each time and frequency sample matches the thermal noise. 
In this case, when the number of antennas $N_{\rm A}$ in a VLBI array is greater than 4 the sensitivity of the bispectrum for that array is greater than of the cross-correlation, which in turn is always greater than the sensitivity of the auto-correlation.
Note that Figure \ref{fig:snr}a is showing the SNR for a single sample interval. This is the equivalent to the integrated SNR for a transient signal, if we take the transient signal as being a single sample wide (i.e. N$_{\rm obs}$=1). This demonstrates that for an array with more than 5 members, when searching for transient signals which do not repeat, the bispectrum method is the most sensitive. 
However, when the SNR is 0.1 (i.e. Figure \ref{fig:snr}b and c) the sensitivity of the bispectrum is much less than the other methods, even that of the autocorrelation. Note that in this case a single transient signal pulse is undetectable for all detection methods as the total SNR per sample is always less than 1. On the other hand, for a pulsar, there are many repeating pulses which can be added together  (i.e. N$_{\rm obs} \gg$1).
% On the other hand for a transient signal across many frequency channels or time samples there are many data samples which can be combined. 
In this case the total detectable signal is the SNR per sample multiplied by the square root of the number of samples. In Figure \ref{fig:snr}c we indicate the integrated SNR which would be obtained for a 6 hour observation of a pulsar such as PSR\,J1745-2900, which would have 5400 pulsations in that time span. This would lead to a detectable signal for both the autocorrelation and cross correlation methods (with the latter having higher sensitivity in the cases when the source position is known). However, this gain is still insufficient to raise the bispectrum SNR to a detectable level. 

We conclude that for transient signals, where one cannot hope for additional samples to improve the integrated SNR, the bispectrum method is the best detection method if your array has more than 4 members. There will be low SNR cases where the auto and cross correlations are more sensitive than the bispectrum method, but in these cases the SNRs will be below detectable levels whatever method is used. 
%Note that this requires that the signal integration is optimal in frequency and time before the formation of the bispectrum. 
For pulsars  where one can combine many single pulses in a coherent timing search the best searching method will be in the phased array mode, if one knows where the pulsar candidate is to be found, or with the sum of the autocorrelations, if the position is unknown.

\section{Observations and Results}
\label{sec:obs}

Pulsar observations in the direction of the Galactic Center below 20\-GHz were predicted to have zero sensitivity for the detection of ms-pulsars ($\tau$=5ms), but this sensitivity rises rapidly for observations at frequencies greater than this cutoff. This is because when the scattering timescales  $\tau_{\rm SC}$ are greater or equal to the pulsar period $\tau$, the pulsed emission is completely washed out, see \citet{macquart_gbt} for details. We have repeated the analytical approach described in \citet{macquart_gbt} and obtain similar results. These are discussed in the following subsections.

\subsection{Expected Frequency of Maximum Fluxes for Scattered Pulsars}
\label{sec:expec}

\begin{figure}
 \includegraphics[width=0.9\textwidth]{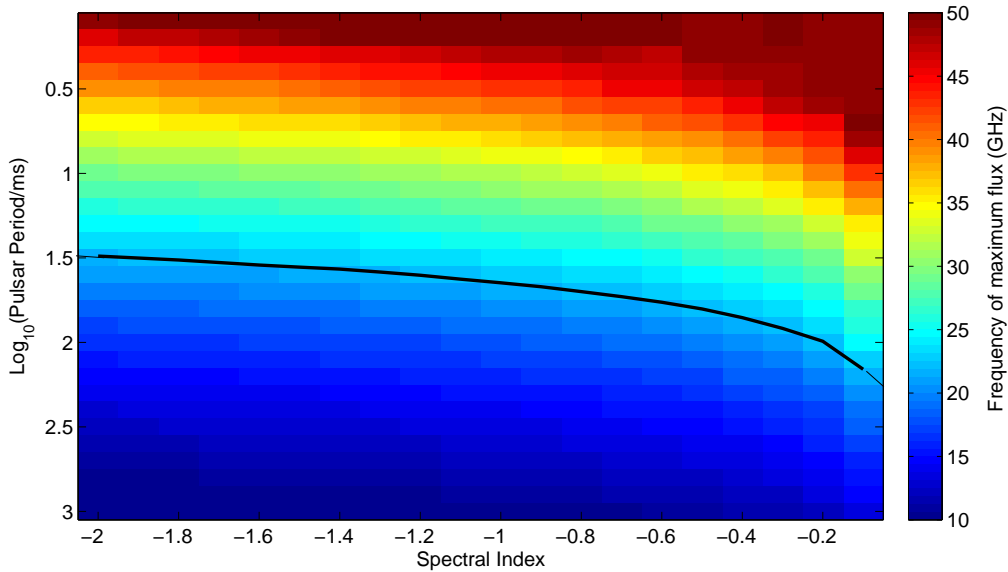}
 \caption{The color scale shows the frequency range between 10--50\,GHz at which the {\it intrinsic} flux would be maximal for pulsars at the Galatic Center plotted against spectral index and pulsar period. The scattering time scale parameters used are those derived in \citet{lazio_98}, and these values were used to smear out a square wave pulsar signal, at the given period, with  a 1\% duty cycle. The losses in the detectable pulsed signal from scattering were combined with the losses expected from the fall in the flux as a function of the spectral index. The frequency at which the maximum signal would be received is plotted, after these losses. No allowance for the variation of the telescope system temperatures is made. A solid line which follows the contour at 22-GHz is overlaid. }
 \label{fig:max-flux}
 \end{figure}

\begin{figure}
 \includegraphics[width=0.9\textwidth]{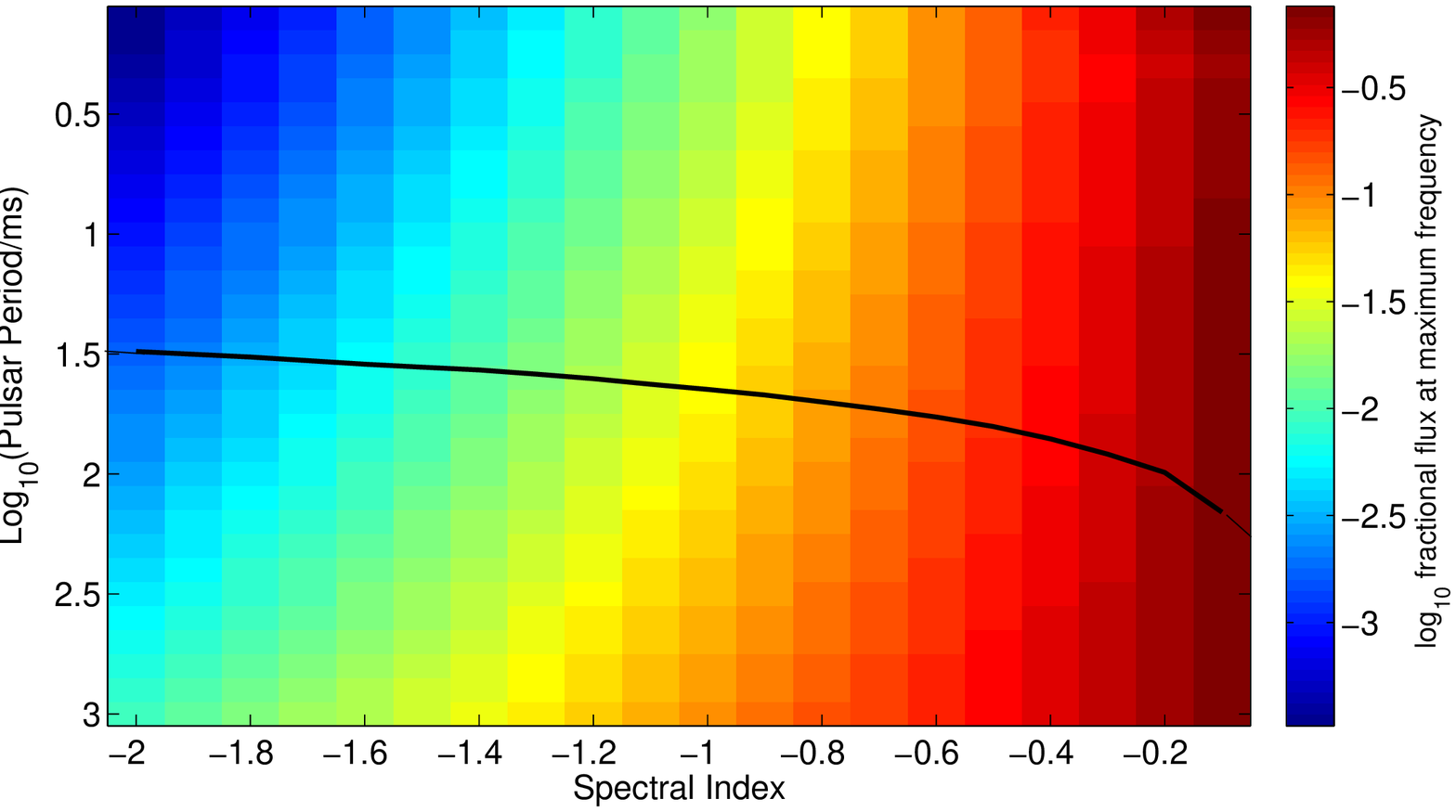}
 \caption{The color scale shows the log$_{10}$ of fraction of flux, compared with that at 1-GHz without any scattering induced losses, that is detectable at the frequency of maximum flux, which is shown in Figure \ref{fig:max-flux}. Note that the short period pulsars are strongly suppressed in strength. Even if we limited ourselves to pulsars with spectral indicies of -1 or flatter we have only a tenth of the signal strength which would have been detected at 1-GHz without any scattering. A solid line for where maximum frequency would be 22-GHz as a function of spectral index and scattering time is overlaid, matching that in Figure \ref{fig:max-flux}.}
 \label{fig:loss-flux}
 \end{figure}

Two dominant effects control the expected signal arriving at the antennas, the fall in flux with increasing frequency because of the negative spectral index and the fall in flux with decreasing frequency because of the scatter smearing of the pulsed signal. The extremely high scattering around Sgr\,A* make the latter effect, which is normally weak compared to the former, dominant over a large frequency range. This is particularly relevant as the pulsars we wish to discover are those which will have the greatest scientific impact. That is those with millisecond periods and, as they are close to the Black Hole, those suffering the highest scattering.
To understand the effects of the scattering on the detectability of these pulsars we have repeated a similar analysis to that taken in \citet{cordes_97} and \citet{macquart_gbt}.
We simulated the effect of the two losses by taking a time series with a square-wave pulsar profile with a 1\% duty cycle. We then convolved this with a one sided exponential filter function with a scattering timescale derived from a scattering screen 100 pc from the Galactic Center, using the data from \citet{lazio_98}. These parameters give an almost identical scattering timescales to those predicted for a dispersion measure of 3000 using the analytical formula in \citet{cordes_03}. We simulated attempting to detect the pulsar by Fourier transforming the scattered and baseline subtracted signal and measuring the peak of the relevant pulsar frequency bin. We obtain more or less indistinguishable results if we average over several harmonic frequency bins. 
We do not find such a sharp cutoff in signal strength as in \citet{macquart_gbt} under our analysis, and this we take to be from the fact we have not included all the effects that were included in that work. However as the analysis can only give an indicative description of the expectations, we have not delved deeper into these effects and limited ourselves to the simplest analysis. 
We plot in Figure \ref{fig:max-flux} the frequency at which we determine the signal would be maximum as a function of Spectral Index and pulsar period. We note that for the short period pulsars (less than 30 milli-seconds), at nearly all spectral indicies, frequencies of 20-GHz and above would be the optimal observing bands. 
Figure \ref{fig:loss-flux} plots the loss in the signal strength at the frequency of that maximum signal compared to the pulsar flux at 1-GHz without scattering. The steep spectrum and short period pulsars suffer losses up to a factor of a thousand, which will make them very difficult to detect. Pulsars with more moderate spectral index, observed around 20-GHz, suffer significant losses but ones which should be within reach of detectability.

\subsection{Phase resolved VLBI imaging}

We began by confirming our understanding of pulsar VLBI processing by imaging a few well known pulsars. 
These were selected based on their spectral index and expected flux at 22-GHz \citep{man_05} and previous detections at higher frequency \citep{kramer_96,kramer_97,morris_97,wiel_00,lohmer_08,eatough_13}.
Experiments k13108a and k13109a were run on the KVN in April 2013 and observed pulsars B0355+54, B0329+54 and B0950+08 in both pointed and in phase referencing modes for several hours each. In May 2013 we had a third epoch with experiment k13131a, which observed B2021+51 in phase referencing mode and also collected six hours of data on the just discovered magnetar PSR\,J1745-2900, for which Sgr\,A* could be used as the in-beam phase reference calibrator. All observations were with a total bandwidth of 256-MHz of Left-Hand Circular Polarisation (LCP), centered at 21.928-GHz.

These first epochs were hampered by the weak flux of the targets and also poor weather conditions. Additionally, the antenna positions still have significant ($\sim$5cm) errors.  Nevertheless, we managed to produce one possible detection of the pulsar B0329+54 when it was phase referenced to J0358+5606. 
%Figure \ref{fig:vlbi-B0329} shows 
We found a 4$\sigma$ detection just to the North East of the error ellipse for the expected pulsar position based on the VLBI measurements of position and proper motion in \citet{brisken_02}. We hope to improve on this result with further observations, particularly with KaVA observations where the better antenna positions and larger number of baselines will aid the detection significantly.

\subsection{Single Dish Filterbank Observations}

We have developed a high time resolution software filterbank which will natively read any VLBI data file recorded in a format supported by the DiFX mark5access libraries. These include the VLBA, Mark4, Mark5, VDIF, K5,  KVN  and VERA formats. The software, {\tt m5fb}, forms a binary file with the averaged power for any feasible spectral and temporal integration. It handles multiple IFs and polarisations. A second program {\tt makeheader} generates a header suitable for introducing the generated filterbank data into the {\tt PRESTO} pipelines. 
%These programs currently assumes that multiple IFs are adjacent, and if these are not more 
All of these programs are bound together with a script {\tt filterbankprep.py}  which extracts much of the configuration information from the VEX file. 

\begin{figure}
  \begin{center}
%    \FigureFile(80mm,80mm){fringe_results.delay.eps}
    \includegraphics[width=0.75\textwidth,angle=-90]{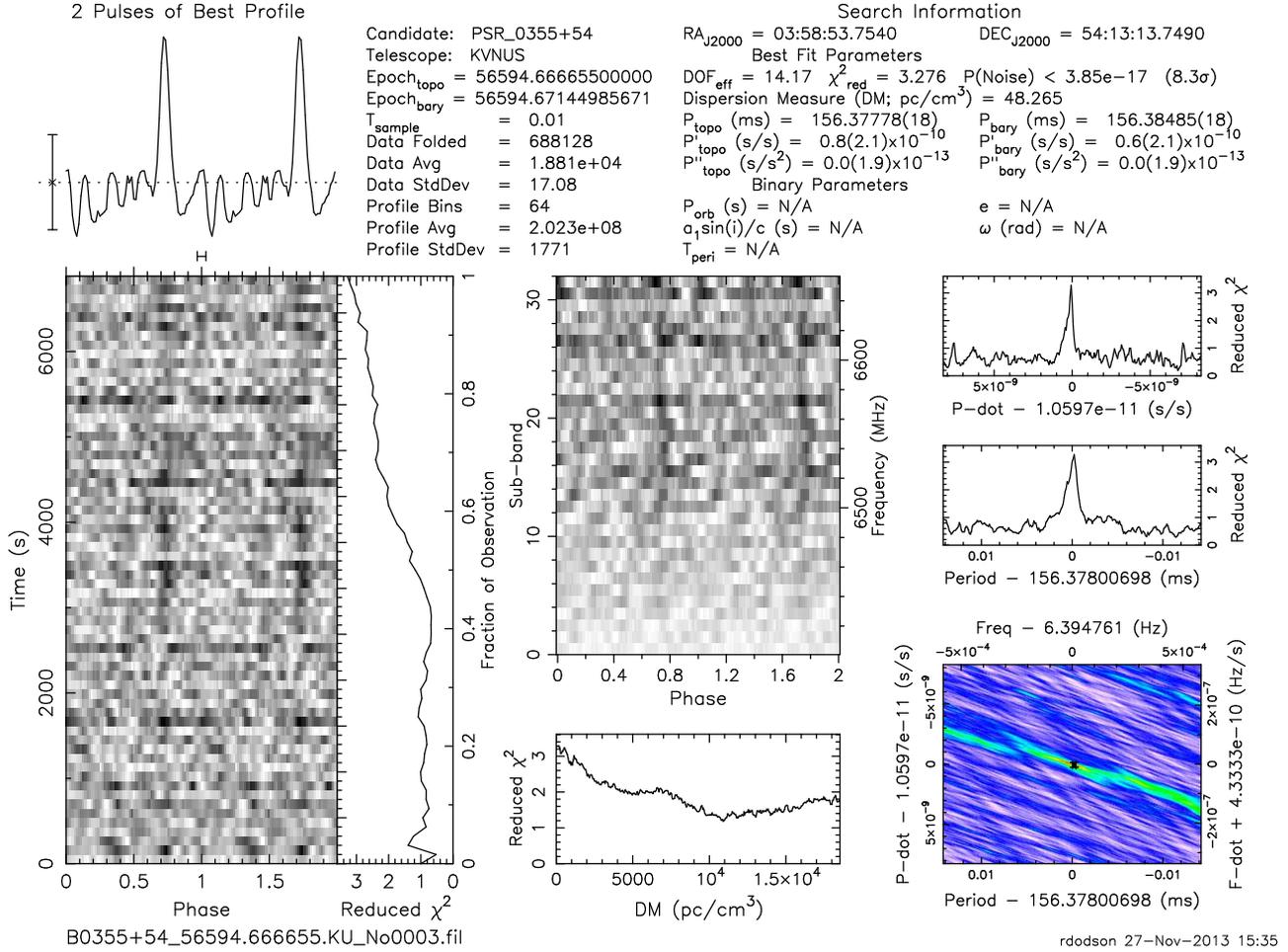}
    %%% \FigureFile(width,height){filename}
  \end{center}
  \caption{Single Dish Filterbank data of well known pulsar B0355+54 at 6.5-GHz, observed at Ulsan and processed via {\tt filterbankprep.py}. The plot itself is generated by {\tt PRESTO}.}
\label{fig:sdish}
\end{figure}

KVN Ulsan has just been equipped with a C-band (6.3 to 8.3-GHz) receiver and our Single Dish filterbank detections have been done with this setup. We recorded 256MHz using the Mark5B recorders, centered at 6.528-GHz, of single dish data on 5 February 2014. 
Two pulsars were detected, B0329+54 and B0355+54 and the latter is shown in Figure \ref{fig:sdish}.  
These were the first pulsar timing detections from the KVN, and the first from our software filterbank. The filterbank is to be found in any distribution of the DiFX software package, as it links to the VLBI data format library, {\tt mark5access}. 

\subsection{Bispectrum time stream}

We have added additional code to DiFX to form the bispectrum during normal correlation operations. This calculates the triple product at every sub-integration, so one has control over the spectral and temporal resolution via the Vex to DiFX control file.
The additional code in the core module loops over all possible triangles and forms the multiples on the averaged spectrally and temporally averaged data, and stores this to a file. Additionally the channels are cumulatively averaged in steps, and integrated in time, before forming further multiples. This allows for more sensitive searches of the integrated values, which span a reduced area on the sky and for longer pulsar periods, in the same correlation pass. 
These complex visibility values are saved during the operation and, as DiFX is an asynchronous correlator, the values need temporal sorting and a smooth complex baseline subtracting after the correlation is complete. The complex baseline represents the slowly changing sky brightness for that baseline and so the average can be over a long timespan. This will provide the upper limit to the maximum timescale of the signal under investigation. 
After these steps, the amplitude of residuals can be saved and passed through the normal pulsar processing tools.

\begin{figure}
  \begin{center}
%    \FigureFile(80mm,80mm){fringe_results.delay.eps}
    \includegraphics[width=0.75\textwidth,angle=-90]{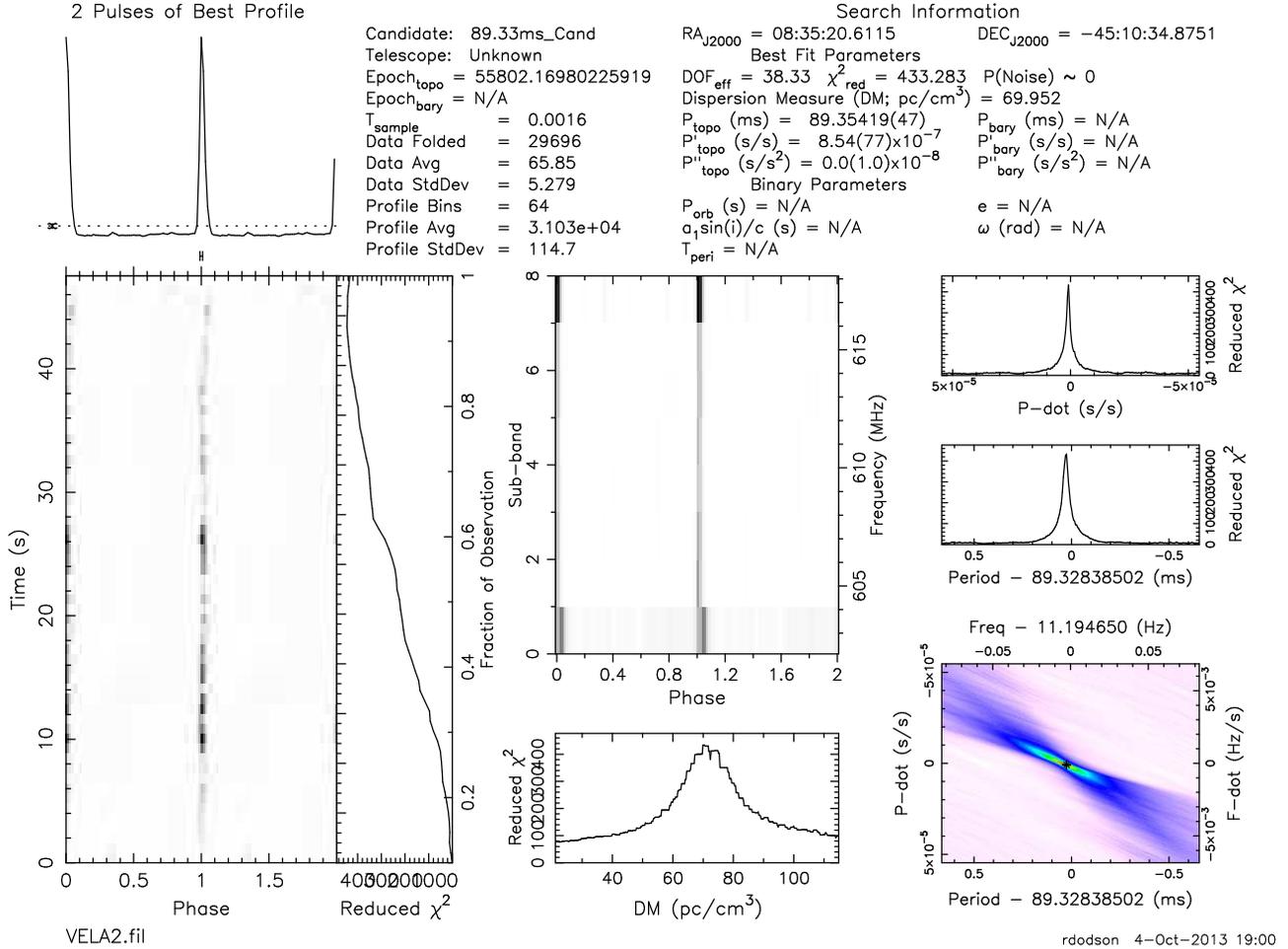}
    %%% \FigureFile(width,height){filename}
  \end{center}
  \caption{The Vela pulsar detection from Bispectrum data formed from GMRT at 610MHz correlated in our modified version of DiFX. The plot itself is generated by {\tt PRESTO}.}
\label{fig:gmrt-vela}
\end{figure}

As we need a signal with high SNR we have demonstrated this method on GMRT observations of the Vela pulsar, which we performed in August 2011. Dual circular polarisation baseband data was recorded from all 30 GMRT antennas at four nearby epochs and two frequencies, 325 and 610MHz. This data has been correlated with the DiFX correlator in pulsar binning mode forming all four Stokes products, so that the extremely bright pulsar flux can be excluded and the PWN can be imaged for comparison with high frequency observations \citep{vela_pwn}. The GMRT correlator \citep{roy_gmrt} is unable to simultaneously provide high time resolution integrations and all polarisation products, therefore this was the only method to obtain such correlator outputs. 
As such this was a very suitable dataset for testing the results from the additional Bispectrum analysis code.
The VLBI-style data collected on Vela was correlated with the phase center several arc-minutes from the pulsar position using our version of DiFX and the bispectrum complex time-stream, integrated to 1.6\,ms and 125\,kHz channel widths, was saved. In total 4060 closure triangles were formed for each channel from each polarisation for each integration.  These data were sorted, summed across all triangles and polarisations, after subtracting a 1 second running average, and the real part extracted and saved in the same format as the data from the high time resolution autocorrelator: 8-bit unsigned integers, with the channels running from high frequency to low frequency for every time sample. This was passed through the same process as before to introduce it to {\tt PRESTO}, as previously described. A header was attached and the pulsar was searched for, using the standard pipelines.
The results from the pulsar search are presented in Figure \ref{fig:gmrt-vela} and the pulsar is clearly detected in every pulse. 

%Note the bleed through for the low frequency edge of the GMRT bandpass is clearly shown with the dispersion measure 

\section{Discussion}
\label{sec:dis}

\begin{table}
\centering
\begin{tabular}{|l|l|}
\hline 
Bispectrum time stream&${\rm SEFD}\ \Omega_{\rm req} / \sqrt{C^{N_{\rm A}}_3} / \sqrt{\Delta \nu \Delta \tau}$\\
Single Dish Filterbank&${\rm SEFD}\ \Omega_{\rm req} / \sqrt{C^{N_{\rm A}}_1} / \sqrt{\Delta \nu \Delta \tau/\Delta {\rm P}}$\\
Phased Array Filterbank&${\rm SEFD}\ \Omega_{\rm req} / \sqrt{C^{N_{\rm A}}_2} / \sqrt{\Delta \nu \Delta \tau/\Delta {\rm P}}$\\
Phase resolved imaging&${\rm SEFD}\ \Omega_{\rm req} / \sqrt{C^{N_{\rm A}}_2} / \sqrt{\Delta \nu \Delta \tau/\Delta {\rm P}}$\\
\hline
\end{tabular}
\caption{Sensitivity formulae for all four observing modes where System Equivalent Flux Density (SEFD) is the typical system temperature in Janskys, $\Omega_{\rm req}$ is the required sensitivity, $C^{N_{\rm A}}_n$ are the number of combinations which can be formed from the $N_{\rm A}$ antennas in that mode, $\Delta \nu$ is the bandwidth, $\Delta \tau$ is the integration time and $\Delta {\rm P}$ is the fraction of the period containing pulsation.}
\label{tab:sen}
\end{table}

\subsection{Bispectrum time stream}

The integration of the Bispectrum filterbank time stream generation during VLBI correlation using DiFX is an important demonstration with a significant spin off for the study of transient emission using arrays. 
It opens a new route to detect time-domain signals in parallel to normal VLBI observing, which hitherto has been limited to incoherent searches in the autocorrelation \citep{vfast}. %, which are necessarily  insensitive

Because the output is formed from the product of three visibility functions the SNR needs to be maximised before the multiplication. The maximum channel width which can be used if one wishes to be sensitive to the whole field of view is small for baselines of thousands of kilometers, so one has the choice of using the shorter baselines or a smaller field of view for maximal sensitivity. We will utilise both of these options, as it is trivial to do so. With 16MHz IFs, each with 128 channels, one MHz channels would be sensitive to the whole Field of View for KVN baselines, whilst the individual channels would be sensitive to the whole Field of View for VERA baselines. If one summed the whole IF for all the KaVA baselines the field of view would be only be about an arcsecond in radius. 
Because of the high frequencies we are targeting in this project we do not need to consider the smearing of the signal due to the high DM of 1700 at the Galactic Center; the DM-induced smearing by this is only 0.3\,ms across the whole 256-MHz bandwidth, at 22-GHz. 

The bispectrum filterbank approach is extremely insensitive to weak signals but very sensitive to stronger ones. Using the expression in Table \ref{tab:sen} and a typical SEFD of 1000 Jy, assuming we average across the entire band and over 1 millisecond we should make a 5$\sigma$ detection of any transient (or single pulse) signal greater than 2\,Jy. Recall that the Lorimer burst was about 30\,Jy and the spectral index is unknown.
% 1000/sqrt(256e6*1e-3)/sqrt(7*6*5/2/3)*5
Therefore, the Bispectrum mode for KaVA offers sensitivity to millisecond-scale signals greater than 2\,Jy, in a Field of View of an arcsecond. This is sufficient to make an interesting target of the Sgr\,A* region where an arcsecond covers the inner most regions where the gravitational fields are intense.

% If this approach was used on ASKAP, with 36 12m antennas with a maximum baseline length of 6km and an observing frequency of 1.4-GHz the channel bandwidth should be less than 3MHz, however dispersion smearing requires that the channel widths are less than this. To limit dispersion smearing to less than 1ms channel widths of 0.2 MHz are required. 

\subsection{Single Dish Filterbank Observations}

The formation of Single Dish Filterbank time streams from VLBI data recordings using the DiFX libraries, which was required for this project to allow the searching for unknown pulsars with the KaVA antennas, is also a significant spin off for the whole astronomical community. 
These datasets are not very sensitive to weak signals and are very prone to interference, however they are the simplest and most reliable wide field of view approach. We are processing all our data in this fashion and search for both pulsations and for transient signals. In a 6 hour observation with 7 antennas at 1Gbps, we should be able to detect a 10$\sigma$ signal with phase averaged pulsation greater than 0.5mJy and a 10\% pulsar period. It is normal to require such a high detection threshold for autocorrelation signals because of their sensitivity to interference.
% 1000/sqrt(256e6*3600*6/0.1)/sqrt(7)*10

\subsection{Phased Array Filterbank Observations}

The plans are well in hand to produce Phased Array Filterbank as an output of software correlation. For software correlation at KASI, we normally use DiFX, as is used at the VLBA, LBA and Bonn, among others. However, for this theme we will use the EVN software correlator SFXC \citep{kettenis_SFXC}, which offers a phased array mode, unlike DiFX. We have demonstrated SFXC correlation on data taken at the KVN and will continue to develop our expertise with this tool. We will correlate our KaVA observations with SFXC as soon as that data is available, to form filterbank time stream outputs on the candidate pulsar sites.
These time streams are sensitive to weak signals and we should be able to make a 5$\sigma$ detection of any pulsation with a phase averaged pulsation flux greater than 0.15mJy and a 10\% pulsar period, with 6 hours of 256MHz recorded bandwidth from the 7 KaVA antennas. %In this case, compared to the one above, we are making no assumptions about the pulsar duty cycle.
%% 1000/sqrt(256e6*3600*6/0.1)/sqrt(7)*10
Once the pulsar period is discovered we will be able to make phase resolved VLBI images of the pulsar using the same data. Note that as the signal can be located spatially as well as temporally our detection cutoffs can be much lower than those for single dish pulsar observations. 

\subsection{Phase resolved VLBI imaging}

Pulsar phase resolved VLBI imaging is a standard DiFX correlation mode, however it was essential to the development of our understanding. The results from this theme are currently just from the KVN and no more than those expected for the standard analysis, so we will not discuss them further. For a pulsar which has a 3\% duty cycle, observed for 6 hours with the seven antennas of KaVA, assuming a typical SEFD of 1000Jy and 256MHz recorded bandwidth, would be expected to have a 5$\sigma$ sensitivity of 0.1mJy. Therefore, for PSR\,1745-2900 for example with a flux of 0.2\,mJy, we could expect a SNR of 10 and an astrometric accuracy on KaVA of 0.1 mas. This is comparable to the best astrometric accuracies achieved with in-beam calibrators at 1.5-GHz by \citet{chatterjee_09}.
%1000/sqrt(256*3600)/sqrt(7*6/2)*sqrt(3/100)*5

\section{Conclusions}
\label{sec:con}

In this work, we present our efforts for pulsar observations with KaVA. Test observations on a few known pulsars have been undertaken with KVN and more are planned with KaVA. We describe four approaches for pulsar observations for observatories with VLBI backends, which have been demonstrated or studied for data processing with the KVN test observations. Each of these approaches will have different signal-to-noise limits and strengths. Therefore, we plan to implement all approaches with the KaVA pulsar pipelines.  We have marginal detections of PSR\,B0329-54 from the KVN pulsar observations with signal-to-noise of a few, but further work is required to confirm this.  Having a greater number of stations with KaVA, we expect to achieve enough sensitivity to confirm and to search for pulsars at 22-GHz or higher.

The comparison of the advantages and relative strengths of the different time domain searching methods can be summarised as: single dish filterbank modes are simple and sensitive to the whole primary beam of the antennas, phased array filterbank modes are significantly more sensitive but are only valid within the very small synthesised beam of the interferometer, whereas the bispectrum time stream is optimal for transient or strong single pulses from pulsars (i.e. non-repeating signals) as they are the most sensitive method for signals detectable in a single integration and they are sensitive to the whole field of view of the correlator. This can be much less than the primary beam of the antenna if the correlator channels are wide, however if the correlator channels are narrow the sensitivity of the bispectrum time stream falls linearly with the number of channels. %

KaVA is capable of pulsar observations and there are good grounds for attempting such observations. One of the most important contributions expected from KaVA pulsar observations will be detection of pulsars in the central region of our Galaxy. The expected sensitivity of different signal processing schemes indicate that KaVA will be able to detect some of the brightest pulsars in the Galactic Center. We plan to perform a deep survey about Sgr\,A* (within a few arcminute$^2$) using the existing KaVA data looking for pulsars in this region. Astrophysically, the Galactic Center is expected to harbor many neutron stars (and pulsars). Although the quantitative estimate is rather uncertain, it is promising that several tens of possible PWN sources have been discovered by {\it Chandra} and that a young magnetar near from Sgr\,A* is observable in the radio bands.  

KaVA pulsar observations should provide ephemerises of pulsars about Sgr\,A* with good precision. This will allow us to map out the gravitational potential in this region and will be useful for follow-up observations at lower frequencies. Over the next a couple of years, we expect KaVA will be able to perform regular observations of pulsars and provide useful information to understand the pulsar population that is bright in the mm/sub-mm frequency range.

\bigskip
\noindent
{\bf Acknowledgements}

\noindent
We are grateful to all members in the KVN and VERA who
help to operate the array. The KVN is a facility operated by the Korea Astronomy and Space Science Institute, VERA is a facility operated by the National Astronomical Observatory of Japan.
RD acknowledges the support of the Korean Ministry of
Science, ICT \& Future Planning Brainpool Fellowship (121S-1-2-0228). 
We acknowledge with gratitude the pulsar solutions provided by Dr Stappers of the University of Manchester. 
The non-standard GMRT observations were performed with the great assistance of Drs Roy and Gupta and many of the support staff of NCRA, India. 

%\bibliography{pulsarproject}

\begin{thebibliography}{50}
\expandafter\ifx\csname natexlab\endcsname\relax\def\natexlab#1{#1}\fi

\bibitem[{{Bates} {et~al.}(2013){Bates}, {Lorimer}, \& {Verbiest}}]{bates_13}
{Bates}, S.~D., {Lorimer}, D.~R., \& {Verbiest}, J.~P.~W. 2013, \mnras, 431,
  1352

\bibitem[{{Bower} {et~al.}(2014){Bower}, {Deller}, {Demorest}, {Brunthaler},
  {Eatough}, {Falcke}, {Kramer}, {Lee}, \& {Spitler}}]{bower_14}
{Bower}, G.~C., {Deller}, A., {Demorest}, P., {et~al.} 2014, \apjl, 780, L2

\bibitem[{{Brisken} {et~al.}(2002){Brisken}, {Benson}, {Goss}, \&
  {Thorsett}}]{brisken_02}
{Brisken}, W.~F., {Benson}, J.~M., {Goss}, W.~M., \& {Thorsett}, S.~E. 2002,
  \apj, 571, 906

\bibitem[{{Camilo} {et~al.}(2006){Camilo}, {Ransom}, {Halpern}, {Reynolds},
  {Helfand}, {Zimmerman}, \& {Sarkissian}}]{camilo_06}
{Camilo}, F., {Ransom}, S.~M., {Halpern}, J.~P., {et~al.} 2006, \nat, 442, 892

\bibitem[{{Chatterjee} {et~al.}(2009){Chatterjee}, {Brisken}, {Vlemmings},
  {Goss}, {Lazio}, {Cordes}, {Thorsett}, {Fomalont}, {Lyne}, \&
  {Kramer}}]{chatterjee_09}
{Chatterjee}, S., {Brisken}, W.~F., {Vlemmings}, W.~H.~T., {et~al.} 2009, \apj,
  698, 250

\bibitem[{{Chennamangalam} \& {Lorimer}(2014)}]{chennamangalam_14}
{Chennamangalam}, J., \& {Lorimer}, D.~R. 2014, \mnras, 440, L86

\bibitem[{{Cole} \& {Pilkington}(1968)}]{cp68}
{Cole}, T.~W., \& {Pilkington}, J.~D.~H. 1968, \nat, 219, 574

\bibitem[{{Cordes} \& {Lazio}(1997)}]{cordes_97}
{Cordes}, J.~M., \& {Lazio}, T.~J.~W. 1997, \apj, 475, 557

\bibitem[{{Cordes} \& {McLaughlin}(2003)}]{cordes_03}
{Cordes}, J.~M., \& {McLaughlin}, M.~A. 2003, \apj, 596, 1142

\bibitem[{Davies \& Large(1970)}]{dl70}
Davies, J.~G., \& Large, M.~I. 1970, \mnras, 149, 301

\bibitem[{{Deller} {et~al.}(2011){Deller}, {Brisken}, {Phillips}, {Morgan},
  {Alef}, {Cappallo}, {Middelberg}, {Romney}, {Rottmann}, {Tingay}, \&
  {Wayth}}]{difx}
{Deller}, A.~T., {Brisken}, W.~F., {Phillips}, C.~J., {et~al.} 2011, \pasp,
  123, 275

\bibitem[{{Deneva} {et~al.}(2009){Deneva}, {Cordes}, \& {Lazio}}]{deneva_09}
{Deneva}, J.~S., {Cordes}, J.~M., \& {Lazio}, T.~J.~W. 2009, \apjl, 702, L177

\bibitem[{{Dodson} {et~al.}(2003){Dodson}, {Lewis}, {McConnell}, \&
  {Deshpande}}]{vela_pwn}
{Dodson}, R., {Lewis}, D., {McConnell}, D., \& {Deshpande}, A.~A. 2003, \mnras,
  343, 116

\bibitem[{{Eatough} {et~al.}(2013{\natexlab{a}}){Eatough}, {Karuppusamy},
  {Champion}, {Keane}, {Lee}, {Kramer}, {Klein}, {Kraus}, {Bassa}, {Lyne},
  {Stappers}, {Spitler}, {Freire}, {Cognard}, {Desvignes}, {Lazarus},
  {Verbiest}, {Brunthaler}, \& {Falcke}}]{ATel_5058}
{Eatough}, R., {Karuppusamy}, R., {Champion}, D., {et~al.} 2013{\natexlab{a}},
  The Astronomer's Telegram, 5058, 1

\bibitem[{{Eatough} {et~al.}(2013{\natexlab{b}}){Eatough}, {Falcke},
  {Karuppusamy}, {Lee}, {Champion}, {Keane}, {Desvignes}, {Schnitzeler},
  {Spitler}, {Kramer}, {Klein}, {Bassa}, {Bower}, {Brunthaler}, {Cognard},
  {Deller}, {Demorest}, {Freire}, {Kraus}, {Lyne}, {Noutsos}, {Stappers}, \&
  {Wex}}]{eatough_13}
{Eatough}, R.~P., {Falcke}, H., {Karuppusamy}, R., {et~al.} 2013{\natexlab{b}},
  \nat, 501, 391

\bibitem[{{Faucher-Gigu{\`e}re} \& {Loeb}(2011)}]{fgl_11}
{Faucher-Gigu{\`e}re}, C.-A., \& {Loeb}, A. 2011, \mnras, 415, 3951

\bibitem[{{Honma} {et~al.}(2003){Honma}, {Fujii}, {Hirota}, {Horiai},
  {Iwadate}, {Jike}, {Kameya}, {Kamohara}, {Kan-Ya}, {Kawaguchi}, {Kobayashi},
  {Kuji}, {Kurayama}, {Manabe}, {Miyaji}, {Nakashima}, {Omodaka}, {Oyama},
  {Sakai}, {Sakakibara}, {Sato}, {Sasao}, {Shibata}, {Shimizu}, {Suda},
  {Tamura}, {Ujihara}, \& {Yoshimura}}]{vera}
{Honma}, M., {Fujii}, T., {Hirota}, T., {et~al.} 2003, \pasj, 55, L57

\bibitem[{{Johnson} {et~al.}(2009){Johnson}, {Dong}, \& {Wang}}]{johnston_09}
{Johnson}, S.~P., {Dong}, H., \& {Wang}, Q.~D. 2009, \mnras, 399, 1429

\bibitem[{{Johnston} {et~al.}(2006){Johnston}, {Kramer}, {Lorimer}, {Lyne},
  {McLaughlin}, {Klein}, \& {Manchester}}]{johnston_06}
{Johnston}, S., {Kramer}, M., {Lorimer}, D.~R., {et~al.} 2006, \mnras, 373, L6

\bibitem[{{Keith} {et~al.}(2011){Keith}, {Johnston}, {Levin}, \&
  {Bailes}}]{keith_11}
{Keith}, M.~J., {Johnston}, S., {Levin}, L., \& {Bailes}, M. 2011, \mnras, 416,
  346

\bibitem[{{Kettenis}(2010)}]{kettenis_SFXC}
{Kettenis}, M. 2010, in 10th European VLBI Network Symposium and EVN Users
  Meeting: VLBI and the New Generation of Radio Arrays

\bibitem[{{Kramer} {et~al.}(1997){Kramer}, {Jessner}, {Doroshenko}, \&
  {Wielebinski}}]{kramer_97}
{Kramer}, M., {Jessner}, A., {Doroshenko}, O., \& {Wielebinski}, R. 1997, \apj,
  488, 364

\bibitem[{{Kramer} {et~al.}(1996){Kramer}, {Xilouris}, {Jessner},
  {Wielebinski}, \& {Timofeev}}]{kramer_96}
{Kramer}, M., {Xilouris}, K.~M., {Jessner}, A., {Wielebinski}, R., \&
  {Timofeev}, M. 1996, \aap, 306, 867

\bibitem[{Large {et~al.}(1968)Large, Vaughan, \& Wielebinski}]{lvw68}
Large, M.~I., Vaughan, A.~E., \& Wielebinski, R. 1968, \nat, 220, 753

\bibitem[{{Law} \& {Bower}(2012)}]{law_12}
{Law}, C.~J., \& {Bower}, G.~C. 2012, \apj, 749, 143

\bibitem[{{Lazio} \& {Cordes}(1998)}]{lazio_98}
{Lazio}, T.~J.~W., \& {Cordes}, J.~M. 1998, \apj, 505, 715

\bibitem[{{Lee} {et~al.}(2014){Lee}, {Petrov}, {Byun}, {Kim}, {Jung}, {Song},
  {Oh}, {Roh}, {Je}, {Wi}, {Sohn}, {Oh}, {Kim}, {Yeom}, {Chung}, {Kang}, {Han},
  {Lee}, {Kim}, {Chung}, {Kim}, {Ryoung Kim}, {Kang}, \& {Cho}}]{kvn}
{Lee}, S.-S., {Petrov}, L., {Byun}, D.-Y., {et~al.} 2014, \aj, 147, 77

\bibitem[{{Liu} {et~al.}(2012){Liu}, {Wex}, {Kramer}, {Cordes}, \&
  {Lazio}}]{liu_12}
{Liu}, K., {Wex}, N., {Kramer}, M., {Cordes}, J.~M., \& {Lazio}, T.~J.~W. 2012,
  \apj, 747, 1

\bibitem[{{L{\"o}hmer} {et~al.}(2008){L{\"o}hmer}, {Jessner},
  {Kramer}, {Wielebinski}, \& {Maron}}]{lohmer_08}
{L{\"o}hmer}, O., {Jessner}, A., {Kramer}, M., {Wielebinski}, R., \& {Maron},
  O. 2008, \aap, 480, 623

\bibitem[{{Lommen} {et~al.}(2007){Lommen}, {Donovan}, {Gwinn}, {Arzoumanian},
  {Harding}, {Strickman}, {Dodson}, {McCulloch}, \& {Moffett}}]{lommen_07}
{Lommen}, A., {Donovan}, J., {Gwinn}, C., {et~al.} 2007, \apj, 657, 436

\bibitem[{{Lu} {et~al.}(2008){Lu}, {Yuan}, \& {Lou}}]{lu_08}
{Lu}, F.~J., {Yuan}, T.~T., \& {Lou}, Y.-Q. 2008, \apj, 673, 915

\bibitem[{{Macquart} {et~al.}(2010){Macquart}, {Kanekar}, {Frail}, \&
  {Ransom}}]{macquart_gbt}
{Macquart}, J.-P., {Kanekar}, N., {Frail}, D.~A., \& {Ransom}, S.~M. 2010,
  \apj, 715, 939

\bibitem[{{Manchester} {et~al.}(2005){Manchester}, {Hobbs}, {Teoh}, \&
  {Hobbs}}]{man_05}
{Manchester}, R.~N., {Hobbs}, G.~B., {Teoh}, A., \& {Hobbs}, M. 2005, \aj, 129,
  1993

\bibitem[{Manchester {et~al.}(1972)Manchester, Taylor, \& Huguenin}]{mth72}
Manchester, R.~N., Taylor, J.~H., \& Huguenin, G.~R. 1972, \nat, 240, 74

\bibitem[{{McGary} {et~al.}(2001){McGary}, {Brisken}, {Fruchter}, {Goss}, \&
  {Thorsett}}]{gating}
{McGary}, R.~S., {Brisken}, W.~F., {Fruchter}, A.~S., {Goss}, W.~M., \&
  {Thorsett}, S.~E. 2001, \aj, 121, 1192

\bibitem[{{Mori} {et~al.}(2013{\natexlab{a}}){Mori}, {Gotthelf}, {Barriere},
  {Hailey}, {Harrison}, {Kaspi}, {Tomsick}, \& {Zhang}}]{ATel_5020}
{Mori}, K., {Gotthelf}, E.~V., {Barriere}, N.~M., {et~al.} 2013{\natexlab{a}},
  The Astronomer's Telegram, 5020, 1

\bibitem[{{Mori} {et~al.}(2013{\natexlab{b}}){Mori}, {Gotthelf}, {Zhang}, {An},
  {Baganoff}, {Barri{\`e}re}, {Beloborodov}, {Boggs}, {Christensen}, {Craig},
  {Dufour}, {Grefenstette}, {Hailey}, {Harrison}, {Hong}, {Kaspi}, {Kennea},
  {Madsen}, {Markwardt}, {Nynka}, {Stern}, {Tomsick}, \& {Zhang}}]{mori_13}
{Mori}, K., {Gotthelf}, E.~V., {Zhang}, S., {et~al.} 2013{\natexlab{b}}, \apjl,
  770, L23

\bibitem[{{Morris} {et~al.}(1997){Morris}, {Kramer}, {Thum}, {Wielebinski},
  {Grewing}, {Penalver}, {Jessner}, {Butin}, \& {Brunswig}}]{morris_97}
{Morris}, D., {Kramer}, M., {Thum}, C., {et~al.} 1997, \aap, 322, L17

\bibitem[{{Muno} {et~al.}(2008){Muno}, {Baganoff}, {Brandt}, {Morris}, \&
  {Starck}}]{muno_08}
{Muno}, M.~P., {Baganoff}, F.~K., {Brandt}, W.~N., {Morris}, M.~R., \&
  {Starck}, J.-L. 2008, \apj, 673, 251

\bibitem[{{Olausen} \& {Kaspi}(2014)}]{olausen_14}
{Olausen}, S.~A., \& {Kaspi}, V.~M. 2014, \apjs, 212, 6

\bibitem[{{Pfahl} \& {Loeb}(2004)}]{pfahl_04}
{Pfahl}, E., \& {Loeb}, A. 2004, \apj, 615, 253

\bibitem[{{Ransom}(2011)}]{presto}
{Ransom}, S. 2011, {PRESTO: PulsaR Exploration and Search TOolkit},
  astrophysics Source Code Library

\bibitem[{{Roy} {et~al.}(2010){Roy}, {Gupta}, {Pen}, {Peterson}, {Kudale}, \&
  {Kodilkar}}]{roy_gmrt}
{Roy}, J., {Gupta}, Y., {Pen}, U.-L., {et~al.} 2010, Experimental Astronomy,
  28, 25

\bibitem[{{Shannon} \& {Johnston}(2013)}]{shannon_13}
{Shannon}, R.~M., \& {Johnston}, S. 2013, \mnras, 435, L29

\bibitem[{{Spitler} {et~al.}(2014){Spitler}, {Lee}, {Eatough}, {Kramer},
  {Karuppusamy}, {Bassa}, {Cognard}, {Desvignes}, {Lyne}, {Stappers}, {Bower},
  {Cordes}, {Champion}, \& {Falcke}}]{spitler_14}
{Spitler}, L.~G., {Lee}, K.~J., {Eatough}, R.~P., {et~al.} 2014, \apjl, 780, L3

\bibitem[{{Thompson} \& {Duncan}(1995)}]{thompson_95}
{Thompson}, C., \& {Duncan}, R.~C. 1995, \mnras, 275, 255

\bibitem[{{Wayth} {et~al.}(2011){Wayth}, {Brisken}, {Deller}, {Majid},
  {Thompson}, {Tingay}, \& {Wagstaff}}]{vfast}
{Wayth}, R.~B., {Brisken}, W.~F., {Deller}, A.~T., {et~al.} 2011, \apj, 735, 97

\bibitem[{{Wharton} {et~al.}(2012){Wharton}, {Chatterjee}, {Cordes}, {Deneva},
  \& {Lazio}}]{wharton_12}
{Wharton}, R.~S., {Chatterjee}, S., {Cordes}, J.~M., {Deneva}, J.~S., \&
  {Lazio}, T.~J.~W. 2012, \apj, 753, 108

\bibitem[{{Wielebinski}(2000)}]{wiel_00}
{Wielebinski}, R. 2000, in Astronomical Society of the Pacific Conference
  Series, Vol. 202, IAU Colloq. 177: Pulsar Astronomy - 2000 and Beyond, ed.
  M.~{Kramer}, N.~{Wex}, \& R.~{Wielebinski}, 205

\end{thebibliography}
%\bibliographystyle{apj}

\end{document}